# Nanosized Vertical Organic Spin-Valves

Robert Göckeritz,[1] Nico Homonnay,[1] Alexander Müller,[1] Tim Richter[1], Bodo Fuhrmann[2], and Georg Schmidt[1,2,a]

[1]Institut für Physik, Martin-Luther-Universität Halle-Wittenberg, 06099 Halle (Saale), Germany
[2]Interdisziplinäres Zentrum für Materialwissenschaften, Martin-Luther-Universität Halle-Wittenberg, 06099 Halle (Saale), Germany
[a]Electronic mail: georg.schmidt@physik.uni-halle.de.

A fabrication process for vertical organic spin-valve devices has been developed which offers the possibility to achieve active device areas of less than 500x500 nm² and is flexible in terms of material choice for the active layers. Characterization of the resulting devices shows a large magnetoresistance of sometimes more than 100 %, however with equally large variation from device to device. Comparison with large-area spin-valves indicates that the magnetoresistance of both, large and small devices most likely originates from tunneling through pinholes and tunneling magnetoresistance.

Organic spin-valve devices are promising candidates for low cost non-volatile electronics, for example for RF-ID tags.[1] Nevertheless, a deeper understanding of the underlying device physics is still necessary, especially of the charge transport mechanisms through the organic material and the origin of the magnetoresistance. In vertical organic spin-valves (OSV) an organic semiconductor (OSC) thin film is sandwiched between two ferromagnetic electrodes and the device is operated by a current flow perpendicular through the layers.[2] Similar to giant magnetoresistance (GMR)[2–5] and tunneling magnetoresistance (TMR)[6–9] the device resistance depends on the relative magnetization of the two electrodes. However, in contrast to GMR in all-metal structures or TMR in metal/oxide structures it is still an open question whether the origin of the magnetoresistance (MR) effects in OSVs published so far is dominated by spin polarized charge transport through the organic spacer layer or by tunneling processes.[10,11] There is, however, a possible approach to gain deeper insight into this problem which takes into account the scaling of the MR effect with device size. In the case of spin injection and charge transport (GMR) the complete device area is expected to contribute to the current path through the OSC. Reducing the size of the device in this case should keep the resistance area product and the relative magnetoresistance constant. Tunneling processes, however, most likely occur in just a small part of the active area.[12] Most probable candidates for tunneling sites would be pinholes at which the thickness of the OSC is reduced to such an extent that tunneling and thus TMR becomes possible.[8–13] Pinholes, however, follow certain statistics resulting in a different resistance change upon downscaling. They vary in size and also exhibit a varying thickness of the remaining OSC layer. Each pinhole thus contributes to the total device characteristics by a different resistance and MR contribution. In order to appreciate the necessary density of pinholes we can use results published by Barraud and coworkers.[14] They create artificial pinholes in an OSC layer using conducting tip atomic force microscopy (AFM) and study the magnetoresistance after the pinhole is filled up with a ferromagnetic metal. The resistance of the pinholes is typically in the range of 100 MΩ while the magnetoresistance can be as high as 300%. In addition, they observe that the resistance strongly varies from pinhole to pinhole and that the MR can even change sign. The resistance of an OSV with pinholes thus consists of the resistance of the OSC layer with the resistance of the pinholes in parallel. The magnetoresistance depends on the individual pinholes present in the device; however, for large-area devices it should always show an average over all pinholes reduced by the parallel resistance of the OSC. An estimate can be done as follows:

We assume that the pinholes exhibit a relative magnetoresistance of 300% and that the parallel resistance of the organic layer is only a tenth of the total resistance of all pinholes in parallel. In this case, we find that the relative magnetoresistance also decreases to approximately one tenth of the original 300%. Although this is a very rough approximation, it becomes clear that for pinholes as investigated by Barraud et al.[14] a density of 0.1-1 pinhole/µm² in a large-area device is enough to yield the standard spin-valve behaviour which is often reported in literature. In reverse, we can conclude that for devices smaller than 1 µm² it should be possible to observe the properties of single pinholes or to be more precise: In sub-micron devices it is most likely to have either no pinhole (very large resistance) or a single pinhole (large MR and medium resistance) while the probability to have more than one pinhole in a single device becomes extremely low. A lot of information can thus be obtained from nanosized OSVs.

Patterning a layer stack for vertical OSV devices in order to define an active area in the nanometer regime is challenging due to the properties of the organic spacer layer which needs to be protected from any kind of damage that may be induced by the fabrication process. Performing optical or electron beam lithography directly on top of the OSC is not possible because of its good solubility in commonly used solvents for this technology and due to a possible damage of the OSC by these chemicals. For most oligomers even contact to water is detrimental.

One approach to avoid the patterning of the layer stack after deposition is the widely used technique of solid shadow-masking during the evaporation processes for defining the device area.[9,15] This is an easily applicable technique and offers full *in situ* fabrication. However, the minimal resulting device dimensions are in the range of a few tens of micrometers and the edges of the active area are actually not well defined due to the finite spacing between mask and sample.

Another existing approach is to use the metallic top electrode to protect the organic film during the lithography steps. This is feasible only for thin organic layers and is reported by Coey *et al.*,[16] where tunneling devices were fabricated with an Tris(8-hydroxyquinolinato)aluminium (Alq3) layer of less than 8 nm thickness and an active area of 5x5 µm² was achieved. The process involves Ar-ion milling for patterning which also raises the question of edge damage of the organic when going to smaller active areas, which may only become detectable when the device size becomes as small as a few 100 nm.

In order to completely avoid standard lithography on top of the organic film Barraud *et al.* use the technique of nanoindentation of a conducting tip AFM into a large-area Alq3 film capped by a resist layer.[14] The resulting nanohole has a diameter of about 20 nm and is filled up by the cobalt top electrode. Because the tunnel current of the indentation tool is used to define the organic thickness, this technique is also limited to a resulting film thickness of the OSC of a few nanometer and thus only to pure TMR devices. Also the influence of the protective resist coating to the underlying Alq3 with a possible contamination of the final layer stack is not discussed. The active area in these devices is determined by the size of the AFM tip resulting in a scalability of device dimensions limited to the lower nanometer range.

In the following we describe a fabrication process which offers the flexibility to scale the device area from millimeter-size down to the nanometer regime while avoiding many of the drawbacks discussed above. Although the active area is defined by standard optical or electron beam lithography, the delicate organic layer does not get in contact with solvents or etching processes and thus cannot be damaged. The thickness of the organic layer can be chosen from zero to a few hundred nanometer. In addition, all metals which are deposited on top of the active organic layer are deposited by magnetron sputtering which avoids possible degradation due to exposure to an electron-beam-evaporator either by X-radiation or electron bombardment. Furthermore, this process offers a large flexibility in the choice of materials for the layer stack of the device.

One of the most studied vertical OSV structure in literature uses a layer stack of $La_{0.7}Sr_{0.3}MnO_3$ (LSMO), followed by Alq3, an additional thin insulating layer and cobalt as the second electrode.[2,4,5,12,13,17,18] We use a similar structure with sputtered MgO as a barrier between the Alq3 and the top electrode. The active device area is defined by a small window in an insulating layer which acts as a kind of nanosized shadow mask.

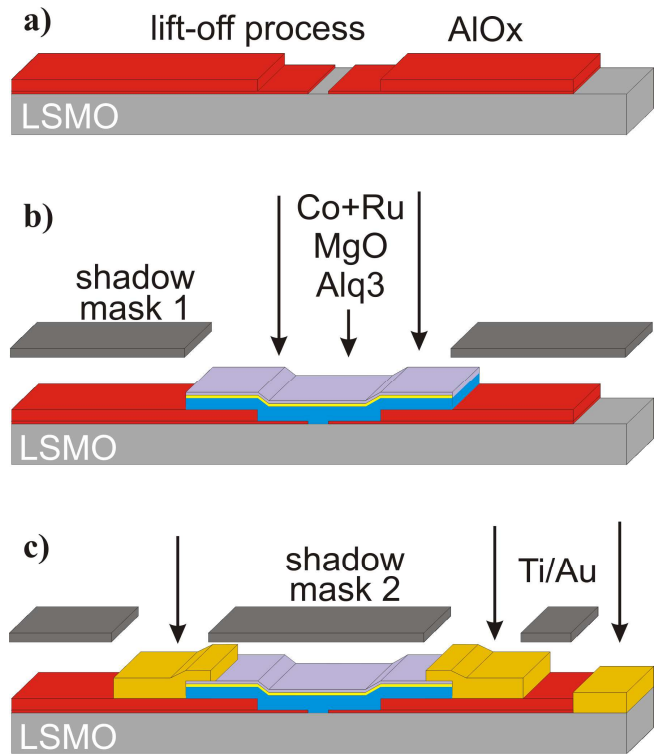

FIG. 1. Schematic representation of the developed fabrication process. The lift-off process of AlOx in (a) can be done using either optical or electron beam lithography. In step (b) all functional layers are deposited using a shadow mask and without breaking the ultra high vacuum. In (c) contact pads are deposited through a second shadow mask after breaking the vacuum.

The ferromagnetic bottom electrode of the device consists of LSMO. Thanks to its good chemical stability in atmosphere and in many solvents it is suitable as substrate for standard lithography steps without surface degradation. The LSMO has a thickness of 20 nm and is deposited on a $SrTiO_3$ substrate by pulsed laser deposition.

The active device area is defined by a two step process. An insulating layer, namely aluminium oxide (AlOx), is deposited by electron beam evaporation (thickness approx. 30 nm) in which a small window is defined by lithography (either electron beam or optical) and lift-off. This window corresponds to the active device area. Subsequently, the step is repeated using a much thicker layer of the same insulator to define a protecting frame around the formerly defined window resulting in the structure of Fig. 1(a). The thin layer allows the deposition of a continuous organic film over the edge of the window while the thick layer is less prone to the formation of pinholes and even withstands the mechanical forces induced by the wire bonding after processing. Outside the frame the LSMO surface is left open to allow for the definition of a back contact. The inset of Fig. 3(c) shows an electron micrograph of an AlOx window with lateral dimensions of approx. 500 nm. The minimum size is only limited by the electron beam lithography based definition of the primary window. Consequently, device dimensions of less than 100 nm are feasible.

Now the sample is ready for *in situ* deposition of the layer stack of Alq3/MgO/Co/Ru [Fig. 1(b)]. For this step a large-area shadow mask is aligned above the pre-patterned substrate. The mask separates multiple devices and prevents organic deposition in the area of the back contact.

The organic semiconductor Alq3 is deposited with a thickness between 12 and 60 nm. Without breaking the UHV the sample is transferred to the sputter chamber and through the same shadow mask MgO is sputtered, followed by the ferromagnetic Co top electrode and a Ru capping layer to prevent the oxidation of the Co. At ambient conditions another shadow mask is aligned and the Ti/Au bonding contacts are deposited with an electron-beam evaporator in UHV [Fig. 1(c)]. This shadow mask defines large-area contacts including the back contact and a large contact pad placed on the insulator where later a wire bond can be deposited. This pad also has some overlap to the Co/Ru. At the same time the shadow mask protects the OSC in the active area inside the AlOx window from any damage by the electron beam evaporation by high energy electrons or X-ray Bremsstrahlung.[19] Finally the sample is bonded for measurements in the cryostat.

Characterization is done in a $^4$He bath cryostat equipped with a 3D vector magnet using low voltage DC excitation.

With the described process samples of 3x5mm² have been processed each of them carrying 7 devices. When electron beam lithography is used, the active device areas are approx. 0.5x0.5 µm² but always one additional large-area control device with about 200x100 µm² is present on the same sample. Using optical lithography active areas ranging from 200x100 µm² down to 5x5 µm² are achieved.

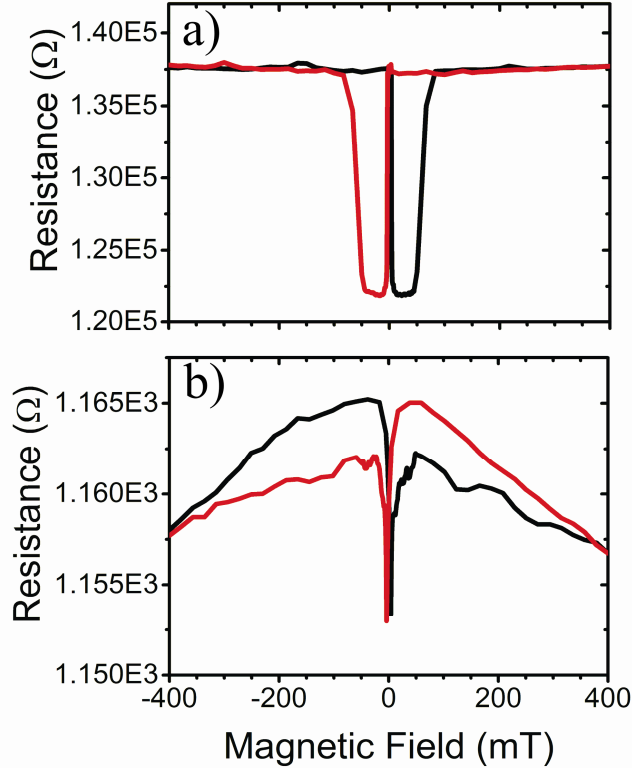

FIG. 2. MR traces of two large-area devices with an active area of 200x100 µm². The Alq3 thickness is 60 nm and 12 nm for device in (a) and (b), respectively.

To first demonstrate the feasibility of the process the large-area reference devices are characterized which were fabricated along with the nanosized OSV structures. The result of an MR measurement of a structure with a large active area of 200x100 µm² is shown in Fig. 2(a). For this device a 60 nm thick organic spacer layer of Alq3 is used which is slightly thinner than the thickness range of several published large-area vertical organic spin-valve devices.[2,12,13,17,18] A spin-valve-like behaviour can be observed with an MR effect of about -11% at a temperature of 4 K and a resistance of about 138 kΩ. A low resistance state occurs between in-plane magnetic fields of ±5 mT and approx. ±50 mT which corresponds to the antiparallel alignment of the two magnetic electrodes.

The reference device with a 12 nm OSC layer exhibits a resistance of approx. 1.1 kΩ which is close to the in-plane resistance of the LSMO contacts. As from simple thickness scaling we would expect a resistance of at least 23 kΩ, this indicates that the organic semiconductor is most likely shorted with a resistance well below 1 kΩ. Furthermore the MR measurement [Fig. 2(b)] has completely different characteristics and only shows an MR of approx. -1% around 0 mT without a pronounced plateau which can be caused by intrinsic MR of the LSMO.

As discussed above, for GMR based devices the resistance should scale inversely with device area (constant resistance area product). Based on the reference sub-micron devices with 60 nm OSC should exhibit a resistance in the range of 1 GΩ while devices with 12 nm OSC should show at least two orders of magnitude less. Nevertheless, the experiments show a completely different behaviour. For scaled down devices with 500x500 nm² active area it is not possible to reliably measure MR effects when the thickness of the organic layer is above 15 nm because the high device resistance exceeded the limits of the measurement setup which is approx. 100 GΩ. For an OSC thickness of approx. 12 nm, however, a variety of different resistances and MR traces are observed in different small area devices, respectively. For at least 60% of the devices the resistance is again beyond the measurement limit. For the remaining devices the resistance is typically in the regime of 10-100 MΩ.

In Fig. 3 MR measurements of four 500 nm-devices from three different samples each with 12 nm Alq3 are shown.

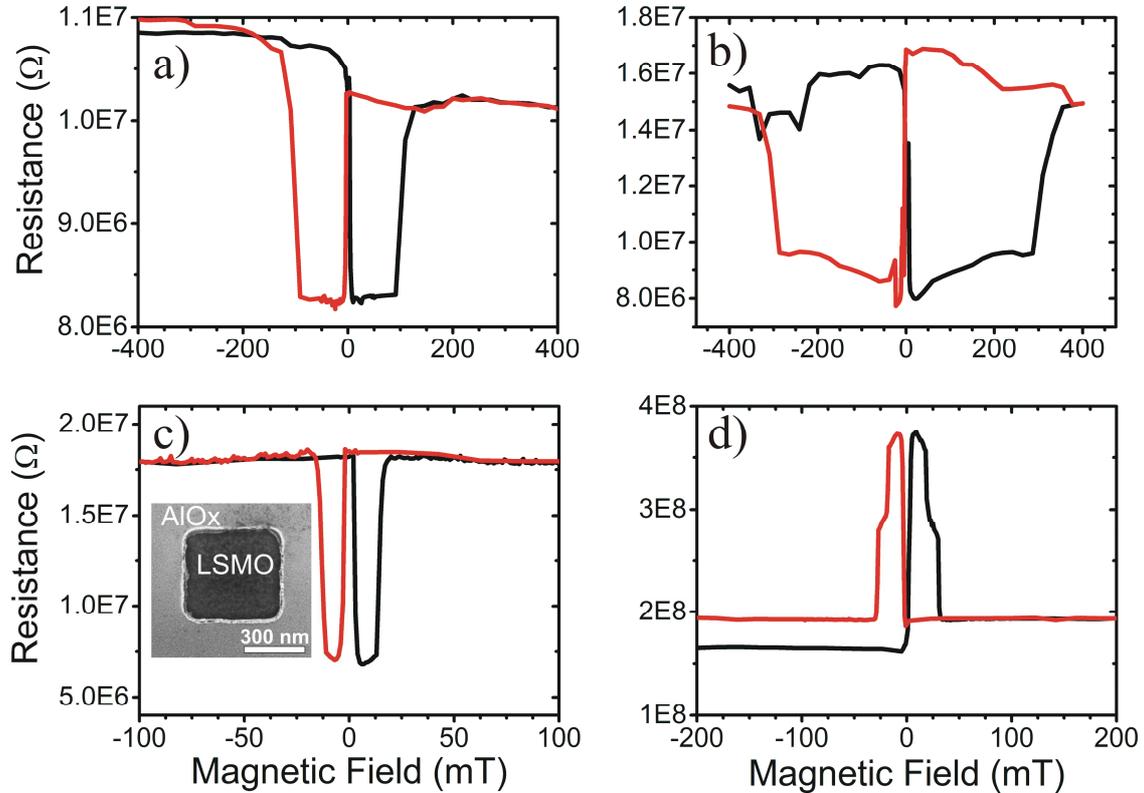

FIG. 3. MR traces of four different devices with an active area of approx. 500x500 nm². Only devices in (b) and (c) are located on the same sample. MR effect varies between -170% and +90%.

The four devices have different total resistance in the saturated state ranging from approx. 10 MΩ for device in Fig. 3(a) up to 200 MΩ for device in Fig. 3(d). Also the magnetoresistance varies in absolute magnitude (20% up to 170%) and sign (minimum -170%, maximum 90%). Furthermore, we observe different magnetic fields for the high field limit of the plateau (between ±50 mT and ±300 mT), which is defined by the switching of the Co while the lower switching field of the LSMO is always at ±5 mT.

This strong variation between different devices can easily be explained, once the picture of the organic spin-valve as a GMR device is abandoned. When we assume transport across pinholes we achieve a consistent picture. As discussed, the devices are then dominated by the individual pinholes and their statistics. Especially for very large devices the probability of a few very low resistance pinholes to occur increases, thus the average device resistance decreases. As all pinhole resistances are in parallel, the lowest resistance pinholes dominate the overall resistance of the device. In the reference device with a 12 nm OSC layer this even results in a short circuit.

For smaller devices the probability to find pinholes in the device area decreases and thus the average resistance of the pinholes present increases, resulting in an increased resistance area product. For the 500 nm devices we thus either see no pinhole at all (beyond measurement limit) or single pinholes with the respective magnetoresistance. Hence even for the 12 nm layer we no longer observe a short circuit but relatively high resistance and sizeable MR. Also the strong variation between different devices and change of sign of the MR is in good agreement with this picture as single pinholes merely probe local properties as seen in the experiments of Barraud et al.[14]

For completeness it should be noted that all devices exhibiting finite MR were investigated in perpendicular magnetic fields[12] and no trace of Hanle effect was detected.

In summary we have fabricated and investigated organic spin-valves with lateral dimensions below one micrometer using a dedicated fabrication process that employs oxide windows as lateral limitations of the devices. The non-linearity in resistance area product upon downscaling and the large variation of the magnetoresistance between different devices indicate that the transport for large-area and nanosized devices is not determined by the organic semiconductor layer itself but rather by pinholes which have a statistical spatial and size distribution. In this case transport and magnetoresistance are mainly dominated by tunneling at the pinhole sites where the OSC is thin enough.

This work was supported by the European Commission within the 7FP project HINTS (Project No. NMP-CT-2006-033370) and by the DFG in the SFB 762. The authors thank Matthias Grünewald for fruitful discussions.